# A High Accuracy and Low Complexity Quality Control Method for Image Compression

Xiao Yan, Zhangxin Gong, Wenqiang Wang, Xiaoyang Zeng, Yibo Fan

*Abstract*—For large-scale still image coding tasks, the processing platform needs to ensure that the coded images meet the quality requirement. Therefore, the quality control algorithms that generate adaptive QP towards a target quality level for image coding are of significant research value. However, the existing quality control methods are limited by low accuracy, excessive computational cost, or temporal information dependence. In this paper, we propose a concise λ domain linear distortion model and an accurate model parameters estimation method based on the original data. Since the model parameters are obtained from the original data, the proposed method is decoupled from the RDO process and can be applied to different image encoders. Experiments show that the proposed quality control algorithm achieves the highest control accuracy and the lowest delay in the literature at the same time. The application of Alibaba's e-commerce platform also shows that the proposed algorithm can significantly reduce the overall bitrate while greatly reducing the bad case ratio.

*Index Terms*—Image Compression, Quality Control, λ domain Distortion Model, Adaptive QP.

## I. Introduction

According to Cisco's research[1], the proportion of video and image data on the Internet has increased from 75% to 82% in the past five years. Applications such as satellites, e-commerce platforms, and social networks usually need to encode a large number of diverse still images and ensure the quality of the resulting images is not lower than a certain threshold. Given the trade-off relationship between image distortion and bitrate, if the image processing platform of these applications can accurately encode all images to a quality level just above the threshold, then the highest overall compression rate can be achieved while the quality requirements are met. Since the mainstream coding standards[2]-[4] use Quantization Parameter (QP) to control the distortion and bit rate of the coded images, the quality control algorithms that generate adaptive QP towards a target distortion for diverse images are of significant research value.

Most of the existing quality control methods aim to minimize the quality fluctuation across the video sequence. They can be divided into Distortion-QP(D-Q) model-based methods[5]-[11], iteration-based methods[12][13] and learning-based methods [15]-[17]. The D-Q model-based methods have a solid theoretical foundation and are the main research direction at present. They attempt to model the relationship between image distortion and QP based on the data distribution, thereby generating an adaptive QP towards the target quality. S. Ma et al.[7] model the distortion as a function of QP and SATD by analyzing the distribution of transform coefficients. C. Seo et al.[8] consider the quadtree coding unit structure and model the probability density function of the transformed coefficients based on a Laplacian function. After that, the Laplacian function-based D-Q and R-Q models are derived to determine the adaptive QP which can minimize the fluctuation of video quality and prevent the overflow and underflow of the bitstream buffer. M. Wang et al.[9] establish a new relationship between the distortion and the Lagrange multiplier with the assumption of constant quality to minimize the distortion variation across video frames at the coding tree unit level. To achieve constant objective reconstruction quality, Q. Cai et al.[10] propose a low complexity preprocessing method to estimate content property without complex RDO and accordingly adjust the Lagrangian multiplier λ. Despite achieving promising results, the above researches on video coding cannot be applied to still image encoding since the parameters of these models are determined based on the coding results of previous frames. To solve this problem, Makarichev et al. [11] proposed an image content-independent D-Q model. They first determine the Upper Bound of Maximal Absolute Deviation (UBMAD) that can be generated by the quantization matrix, and then establish the correlation between the image distortion and the UBMAD. However, the resulting coding distortion can only be controlled in a wide range since the proposed D-Q model does not take the image content into consideration.

The most straightforward quality control method is to iteratively encode the image with different QPs until the quality target is met. Fouzi et al. [12] iteratively perform transformation, quantization, inverse quantization, and inverse transformation process to control the image quality. Similarly, Miaou and Chen [13] firstly determine the distortion target for local image region through a search strategy, and then iteratively encode the image blocks to the distortion target. However, the iteration-based methods introduce a huge computational delay. Furthermore, L.

This work was supported in part by Alibaba Innovative Research (AIR) Program, in part by Shanghai Science and Technology Committee (STCSM) under Grant 19511104300, in part by National Natural Science Foundation of China under Grant 61674041, in part by the Innovation Program of Shanghai Municipal Education Commission, in part by the Fudan University-CIOMP Joint Fund (FC2019-001) (Corresponding author: Xiao Yan).

Xiao Yan is with Electronics Engineering Department, Xi'an University of Posts and Telecommunications, Xi'an 710121, China (xiao_yanxd@163.com)

Zhangxin Gong, Yibo Fan, Xiaoyang Zeng are with the State Key Laboratory of ASIC and System, Fudan University, Shanghai 200433, China ({21112020131, fanyibo, xyzeng}@fudan.edu.cn;).

Wenqiang Wang is with Alibaba Group, Hangzhou, Zhejiang 311121, China (channing.wwq@alibaba-inc.com).

Li et al. [14] show that aligning all local distortions to a fixed quality level will degrade the global compression performance.

Compared with the above traditional methods, many studies also explore the learning-based quality control methods. X. Pan and Z. Chen [15] use a ν-Support Vector Regression (SVR) method to model the rate-distortion features of different video sequences and guide the quantization process through the proposed model and temporal information. M. Santamaria et al. [16] use a convolutional neural network to predict the rate-distortion results at different QPs for intra frames. Since only the intra-frame information is used, this method can be directly used for image coding. However, tests show that this method has a large computational delay. Hou et al.[17] extract 23 feature vectors (3 temporal features, 20 spatial features) from video sequences, and then use support vector regression to model the relationship between the feature vectors, target PSNRs, and optimal QPs. The resulting regression function is used to generate the adaptive QP towards the target quality.

Through the above investigation, it can be seen that there is no low-computation, high-accuracy, and temporal information-independent quality control algorithm for still image coding in the literature. Therefore, the coding platforms usually encode all images with a conservative fixed QP to ensure the complex images meet the quality threshold, making the simple images lie in an unnecessary high-quality range and waste lots of bits.

To solve this problem, this paper derives a linear D-$\lambda$ model based on the $\lambda$-domain distortion model[14] and then proposes a novel mothed that estimates the image content-related model parameters through the distortion characteristics of the original pixels. In this case, for any image to be compressed, we can accurately build its D-$\lambda$ model without actually encoding it, and then determine the Lagrangian multiplier $\lambda$ towards the target quality. After that, the adaptive QP towards the target quality can be determined according to the $\lambda$ and the recommended QP-$\lambda$ relationship[18] provided by the standard organization. Since the image content-related parameters are estimated from the original data, the computational load of the proposed quality control algorithm is significantly lower than that of the existing methods. In addition, the proposed algorithm is completely independent of the implementation of the image encoder and is feasible for different encoders. Experiments show the proposed algorithm achieves the highest control accuracy and the lowest computational delay at the same time in the literature.

## II. Method

### A. The linear D-$\lambda$ model

Li et al. [19] show that ideally all modes in the encoding process should be determined by the rate-distortion cost, and the Lagrange multiplier λ is the most basic hyper-parameter. Therefore, they establish the R-λ model and QP-λ model based on the statistical data[18]. Based on these two models, the video coding task firstly determines an appropriate λ towards the target bitrate $R_t$, and then obtain the QP according to the λ ($R_t$->λ ->QP). This is the basic idea of the current mainstream rate control algorithms.

As the dual problem of rate control, the proposed quality control algorithm attempts to use a λ domain distortion model (D-λ model) to determine the λ towards the target distortion $D_t$, and then obtain the corresponding QP ($D_t$->λ->QP) based on the QP-λ model. In fact, Li et al. [14] have already derived the following D-λ model:

$$D = \alpha \cdot \lambda^{\beta} \quad (1)$$

where D is represented by MSE, $\alpha$ and β are the image content-related model parameters. In the video coding, $\alpha$ and β are iteratively determined by the encoding results of previous frames. Due to the lack of temporal information, it is difficult to determine $\alpha$ and β for still images. Therefore, the main problem of this paper is to accurately estimate the image content-related model parameters without encoding the image.

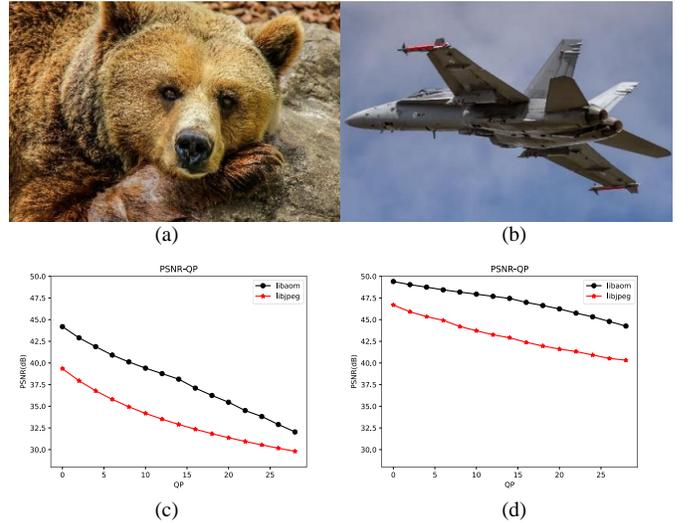

Fig. 1. D-Q curves for different encoders. (c) and (d) show the relationship between PSNR and QP for (a) and (b) with different encoders, respectively.

Similar to the related works[8]-[10], this paper uses PSNR as the distortion metric. According to the relationship between MSE and PSNR, the D-λ model in Eq. (1) can be transformed into the following PSNR-based form:

$$PSNR = a \cdot \log_{10}(\lambda) + b \quad (2)$$

Where $a$ and $b$ are the image content-related parameters. It can be seen that the PSNR of the encoded image is a linear function of the Lagrange multiplier $\lambda$ in logarithmic domain. Compare with Eq. (1), the linear model is easier to fit and calculate.

The parameters in Eq. (2) can be estimated by two or more (PSNR, $\log_{10}(\lambda)$) pairs, and then the $\lambda$ for any target PSNR can be quickly determined. However, due to the high complexity of current encoders, repeatedly encoding the image with different $\lambda$ incurs a huge computational cost. Therefore, the key to building the linear D-λ model becomes to estimate several (PSNR, $\log_{10}(\lambda)$) pairs with a low computation load.

### B. The estimation of image content-related parameters

By analyzing a large amount of data, we find that the D-Q curves of the same image generated by different encoders have similar trends. Fig.1 shows two images and their corresponding D-Q curves generated by JPEG encoder libjpeg [20] and AV1 reference model libaom[21]. It should be noted that the image quality of libjpeg increases as the QP increases, while the image quality of libaom decreases as the QP increases. To highlight

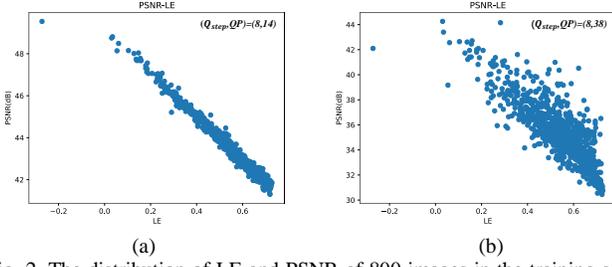

Fig. 2. The distribution of LE and PSNR of 800 images in the training set of DIV2K with different ($Q_{step}$, QP). It can be seen that when ($Q_{step}$, QP)=(8,14), LE of the original data and PSNR generated by libaom show a significant linear relationship while when ($Q_{step}$, QP)=(8,38), their correlation is weak.

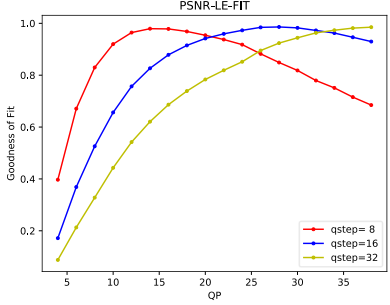

Fig. 3. The goodness of linear fitting with different ($Q_{step}$, QP). The best linear fitting for libaom can be obtained when ($Q_{step}$,QP) = (8,14), (16,28), (32,38).

the correlation between the two D-Q curves, we perform the following transformations on the two QPs:

$$\begin{cases} QP_{jpg,i} = \max(QP_{jpg}^{ori}) - QP_{jpg,i}^{ori} \\ QP_{av1,i} = QP_{av1,i}^{ori} - \min(QP_{av1}^{ori}) \end{cases} \quad (3)$$

where $i$ is the index of the QP, $QP_{jpg,i}^{ori}$ and $QP_{av1,i}^{ori}$ are the i-th original QP of libjpeg and libaom respectively. After this transformation, the QPs will both start from 0, and the PSNR will both decrease as the QP increases.

Comparing the D-Q curves in Fig.1 (c) and (d), it can be seen that images with higher starting PSNR and smaller slope with libjpeg also show a similar result with libaom and vice versa. The root cause for this phenomenon is that the distortion of JPEG images is determined by the distribution of original data in the DCT domain, while the distortion of encoders such as HEVC and AV1 is determined by the distribution of residual data in complex transformation domains such as DCT, DST, and ADST. A detailed derivation of the relationship between distortion and coefficient distribution can be found in [5]-[7]. Despite the complex partition, prediction, and transformation processes, the residual data is rooted in the original pixel data and therefore these D-Q curves show implicit correlations.

Inspired by this observation, this paper tries to estimate the (PSNR, $\log_{10}(\lambda)$) pairs of complex encoders from the original data. Similar to the quantization process of JPEG, we define the distortion of original data as:

$$\begin{cases} MSE = \sum_{n=0}^{N-1} \sum_{i=0}^{63} \left[\left(Coef_i + \frac{Q_{step}}{2}\right) \% Q_{step}\right]^2 / N / 64 \\ LE = \log_{10}(MSE) \end{cases} \quad (4)$$

Where $Coef_i$ is the DCT coefficient of the original data after 8x8 partition, $Q_{step}$ is the quantization step, $i$ is the frequency index, N is the number of the 8x8 blocks, LE is the quantization distortion of the original data in the logarithmic domain.

Taking AV1 as an example, as shown in Fig.2, we find that LE shows a significant linear relationship with the PSNR generated by libaom with appropriate $Q_{step}$ and QP. For the convenience of calculation, we set $Q_{step}$ as 8, 16, 32 and use the goodness of linear fit as the criteria. As shown in Fig.3, when ($Q_{step}$, QP) takes (8, 14), (16, 28), (32, 38), the optimal linear fitting of the LE-PSNR model can be obtained as follows:

$$\begin{cases} PSNR = -10.27 \cdot LE + 49.11, (Q_{step}, QP) = (8,14) \\ PSNR = -11.18 \cdot LE + 50.10, (Q_{step}, QP) = (16,28) \\ PSNR = -11.21 \cdot LE + 51.39, (Q_{step}, QP) = (32,38) \end{cases} \quad (5)$$

It is worth noting that the parameters in Eq.(5) are independent of the images. For any new image, we can calculate its LE with $Q_{step}$=8, 16, and 32 based on the original data, and then estimate its PSNR generated by libaom with QP=14, 28, and 38 according to Eq.(5), finally get three (PSNR, $\log_{10}(\lambda)$) pairs through the recommended QP-$\lambda$ model. After that, the image content-related parameters a and b in Eq.(2) can be easily determined by another linear fitting. It is worth noting that we use $Q_{step}$=8, 16, 32 just to simplify the mod operation in Eq.(4), any other $Q_{step}$ can also be used as long as the goodness of fit in Eq.(5) is high enough. In addition, we can also take more ($Q_{step}$, QP) and (PSNR, $\log_{10}(\lambda)$) samples to improve the accuracy of the estimated D-$\lambda$ model when it is necessary.

*C. The quality control algorithm for still image coding*

For crystal clarity, this section is used to restate the proposed quality control algorithm. This algorithm consists of training and inference phases. The training process is based on a large amount of experimental data to fit the relationship between the original distortion LE and the PSNR generated by the complex encoders, and the resulting linear relationships between LE and PSNR with three ($Q_{step}, QP$) are shown in Eq.(5), which is independent of the image content. For a new image to be encoded, the inference process first calculates its LE value with three $Q_{step}$ and estimates three (PSNR, QP) pairs generated by complex encoders according to Eq.(5). After that, three pairs of (PSNR, $\log_{10}(\lambda)$) can be determined by the QP-$\lambda$ model and finally, the linear distortion model shown in Eq.(2) can be fitted.

It should be noted that the proposed algorithm only needs to perform one DCT transformation, calculate three LE, three multiply-add operations, and one linear interpolation to build the D-$\lambda$ model shown in Eq.(2) for a new image, resulting in an extremely low computational load.

## III. EXPERIMENTS AND RESULTS

*A. Image set and experiment conditions*

In order to demonstrate the universality of the proposed algorithm, this paper conducts experiments on H.265 encoder HM 16.7, X265 3.4 and AV1 encoder libaom 3.1.0 respectively. These encoders are configured as single-pass, fixed-QP mode to encode the still images. DIV2K is a high-quality imageset with a resolution of approximately 1920x1080 provided by the NTIRE 2017 Image Superresolution Challenge, including 800 images in the training set, 100 images in the validation set, and

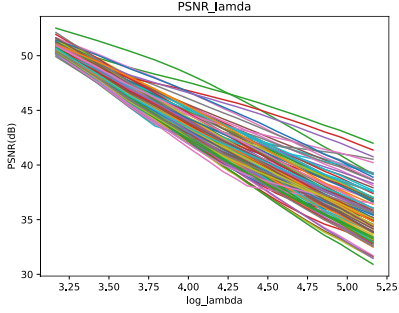

Fig. 4. The linear relationship between PSNR and $log10(\lambda)$.

100 images in the test set[22]. The training set is used to fit the PSNR-LE model while the validation set, test set, and Kodak set are used to test the quality control accuracy. All of the tests are executed on an Intel Xeon @3.20GHz CPU. Similar to [10], the metrics to measure the performances of the quality control methods are defined as:

$$\Delta\% = \frac{|PSNR_m - PSNR_t|}{PSNR_t} \cdot 100\% \quad (6)$$

$$\sigma = \frac{1}{N-1} \cdot \sum_{i=0}^{N-1}(PSNR_t - PSNR_i)^2 \quad (7)$$

Where $PSNR_m$ is the mean PSNR of the test images, $PSNR_t$ is the target PSNR of the quality control algorithm, $\Delta\%$ and $\sigma$ are the control difference (Diff.) and the quality variance (Vari.) of the test images respectively. Similar to Hou et al.[17], this paper conducts tests on the luma component since the experiments show that the distortion of the chroma is much lower than that of the luma with the same QP and the latter basically determines the overall distortion of the image.

### B. The PSNR-$log_{10}(\lambda)$ model

To demonstrate the linear relationship between PSNR and $log10(\lambda)$, we randomly select 100 images in the training set and encoded them with libaom and different $\lambda$. The relationship between the resulting PSNR and $log10(\lambda)$ is shown in Fig.4. It can be seen that the results are consistent with the linear relationship shown in Eq.(2) and different images correspond to different parameters a and b.

### C. The results of the quality control algorithm

In the training phase, we use the image encoders to compress the training images in DIV2K set with QPs from 4 to 40, record the PSNRs, QPs, and $\lambda$s, and calculate the LE of each image with $Q_{step}$= [8, 16, 32] at the same time. After that, ($Q_{step}$, QP) pairs with the highest goodness of fit and the corresponding PSNR-LE model can be determined. The results of libaom, x265, and HM are shown in Table I.

According to the resulting PSNR-LE models, we evaluate the control difference and quality variance of the proposed quality control algorithm based on the validation set, test set, and Kodak set. To be more convincing, we set the target PSNR from 35dB to 45dB in 1dB steps, and take the average $\Delta\%$ and $\sigma$ as the final results. As shown in Table II, the results of existing works [7], [15]-[17] are used for comparison. It should be noted that their results are tested on medical video sequences with very stable temporal characteristics [17], while the proposed quality control algorithm is tested on the independent still images and achieves a significantly lower control difference in the absence of temporal information. In addition, the operation delay of the proposed algorithm is also at least two orders of magnitude lower than other algorithms.

TABLE I
THE PARAMETERS TO BUILD THE PSNR-LE MODEL

| Encoders | libaom | x265 | HM |
|---|---|---|---|
| ($Q_{step}$,QP) | (8,14), (16,28), (32,38) | (8,24), (16,30), (32,36) | (8,22), (16,28), (32,34) |
| Goodness of Fit | 0.9794, 0.9860, 0.9851 | 0.9855, 0.9923, 0.9920 | 0.9764, 0.9897, 0.9915 |
| Parameters of PSNR-LE Model | (-10.27,49.11) (-11.18,50.10) (-11.21,51.39) | (-10.17,49.12) (-10.82,50.12) (-11.19,51.23) | (-10.76,48.65), (-11.15,49.64), (-11.43,50.70) |

TABLE II
COMPARISON OF DIFFERENT QUALITY CONTROL METHODS ON IMAGE SET

| Methods | [7] | [15] | [16] | [17] | Ours | | |
|---|---|---|---|---|---|---|---|
| Encoders | HM | HM | HM | HM | libaom | x265 | HM |
| Diff. (%) | 5.3 | 4.2 | 4.1 | 2.6 | 0.23 | 0.18 | 0.23 |
| Vari. | - | - | - | - | 0.13 | 0.11 | 0.15 |
| Time(ms) | 18648 | 2167 | 20286 | 3341 | 29 | | |

TABLE III
COMPARISON OF DIFFERENT QUALITY CONTROL METHODS ON CLASS E

| Seq. | Target PSNR | [10]-CQ Diff.(%) | [10]-CQ Vari. | [10]-CQL Diff.(%) | [10]-CQL Vari. | Ours Diff.(%) | Ours Vari. |
|---|---|---|---|---|---|---|---|
| Four People | 43.83 | 0.43 | 0.0037 | 0.59 | 0.0050 | 0.06 | 0.0006 |
| | 41.28 | 0.39 | 0.0137 | 0.62 | 0.0022 | 1.15 | 0.0012 |
| | 38.39 | 0.80 | 0.0058 | 1.40 | 0.0049 | 0.57 | 0.0673 |
| | 35.34 | 0.97 | 0.0065 | 1.63 | 0.0113 | 0.23 | 0.0051 |
| Johnny | 44.10 | 0.16 | 0.0084 | 0.32 | 0.0022 | 0.12 | 0.0021 |
| | 41.98 | 0.18 | 0.0026 | 0.44 | 0.0056 | 0.47 | 0.0221 |
| | 39.71 | 0.97 | 0.0090 | 0.91 | 0.0046 | 0.71 | 0.0043 |
| | 37.21 | 1.36 | 0.0068 | 1.39 | 0.0060 | 0.64 | 0.0045 |
| Kristen andSara | 44.49 | 0.40 | 0.0023 | 0.49 | 0.0024 | 0.16 | 0.0028 |
| | 42.17 | 0.21 | 0.0044 | 0.60 | 0.0039 | 0.85 | 0.0167 |
| | 39.24 | 0.76 | 0.0200 | 0.72 | 0.0419 | 0.59 | 0.0138 |
| | 36.33 | 1.00 | 0.0260 | 1.00 | 0.0201 | 0.12 | 0.0406 |

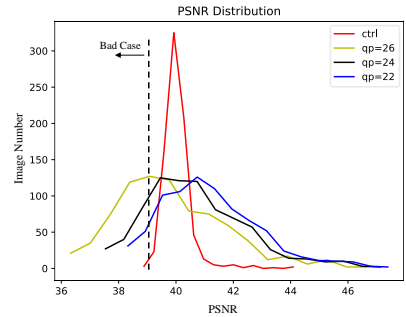

Fig. 5. The quality distribution of 1000 images in Alibaba's e-commerce platform with different encoding mode.

The method proposed in Q. Cai et al. [10] is the state-of-the-art quality control algorithm for video coding. To compare with it, we further build the LE-PSNR model on each frame of Class B sequences and test the proposed quality control algorithm on the frames of Class E sequences in all intra mode. The results are shown in Table III and the best results are highlighted. It should be noted that the target PSNR in [10] depends on the coding result of the first frame of the sequence, while that of the proposed algorithm can be arbitrarily specified. However, compared with [10], the proposed algorithm lacks the ability of rate control since it is not necessary for still image compression. In terms of the encoding delays, the increase of Q. Cai et al. [10] is 0.51% while that of this paper is just about 0.20%.

TABLE IV
COMPARISON OF DIFFERENT IMAGE ENCODER

| Methods | QP=26 | QP=24 | QP=22 | Ours |
|---|---|---|---|---|
| Mean PSNR (dB) | 39.8 | 40.66 | 41.17 | 40.006 |
| Mean Bitrate (bpp) | 0.717 | 0.811 | 0.873 | 0.810 |
| Bad Case Ratio (%) | 36.6 | 17.1 | 7.3 | 1.6 |

To highlight the practical value of the proposed algorithm, we use libaom to encode 1000 images in Alibaba's e-commerce platform in fixed QP and quality control mode and show the quality distribution in Fig. 5. When we set the target quality as 40dB and regard the images with PSNR lower than 39dB as bad cases, the mean PSNR, mean bitrate and bad case ratio results of QP=26, 24, 22 in fixed QP mode and the quality control mode are shown in Table IV. It can be seen that the fixed QP mode has to improve the quality of all images to reduce the bad case, while the proposed quality control algorithm drives the images to the target quality, therefore greatly reducing the bad cases while saving the bitrate by 7.22% compared to QP=22.

## IV. CONCLUSION

This paper presents the first practical quality control algorithm for still image coding. According to the experiments, it can be seen that the proposed algorithm achieves the highest control accuracy and the lowest computational delay at the same time in the literature. In addition, the proposed quality control algorithm is independent of the encoder and can be easily applied to different encoders and coding standards. The application results from Alibaba's e-commerce platform show that the algorithm can significantly reduce the overall bitrate while greatly reducing the bad case ratio.